# TSTNN: TWO-STAGE TRANSFORMER BASED NEURAL NETWORK FOR SPEECH ENHANCEMENT IN THE TIME DOMAIN

*Kai Wang, Bengbeng He, Wei-Ping Zhu*

Department of Electrical and Computer Engineering, Concordia University, Montreal, Canada

## ABSTRACT

In this paper, we propose a transformer-based architecture, called two-stage transformer neural network (TSTNN) for end-to-end speech denoising in the time domain. The proposed model is composed of an encoder, a two-stage transformer module (TSTM), a masking module and a decoder. The encoder maps input noisy speech into feature representation. The TSTM exploits four stacked two-stage transformer blocks to efficiently extract local and global information from the encoder output stage by stage. The masking module creates a mask which will be multiplied with the encoder output. Finally, the decoder uses the masked encoder feature to reconstruct the enhanced speech. Experimental results on the benchmark dataset show that the TSTNN outperforms most state-of-the-art models in time or frequency domain while having significantly lower model complexity.

## 1. INTRODUCTION

Speech enhancement as an indispensable front-end task in many speech processing related applications, such as automatic speech recognition, hearing aid, telecommunication and so on, has been widely studied over the past decades, especially in recent years when deep learning emerges as a powerful tool to develop various data-driven approaches for solving traditional estimation problems with or without supervision.

Most of current deep learning architectures for speech enhancement are implemented in time-frequency (T-F) domain, such as convolutional neural network (CNN) and recurrent neural network (RNN). By using short-time Fourier transform (STFT), those methods usually treat the spectral magnitude as training target. The phase of noisy speech is utilized along with the enhanced speech magnitude to reconstruct the time-domain signal with inverse short-time Fourier transform (iSTFT). Despite some inspiring results achieved [1-4], there are still two main limitations in T-F domain methods. First, Fourier transform operation is an additional overhead which hinders fast speech denoising. Second, the noisy phase information is usually ignored during denoising process. However, the phase information has been proved important for enhancing the speech quality [5]. Some studies have considered both magnitude and phase simultaneously during the training stage to achieve better enhancement results [6].

Recently, various works have directly estimated the clean waveform from noisy raw data in time domain [7-11]. Many researchers have investigated encoder-decoder frameworks based on the CNN or RNN. For modeling long-range sequence like speech, CNN requires more convolutional layers to enlarge receptive field.

The dilated convolutional neural network has been proposed for processing the long-term temporal sequence [12]. Additionally, RNN such as long short-term memory (LSTM) and gated recurrent units (GRU), are commonly used in modeling long-term sequence with order information. However, the drawback of RNN based models is that they cannot perform parallel processing, thus leading to high computation complexity. Although some improvements can be achieved by adding temporal convolutional network (TCN) blocks [10] or LSTM layers between encoder and decoder for further extracting high-level features and enlarging receptive fields [11], the contextual information of speech is often ignored, which restricts the denoising performance. Fortunately, transformer neural network can resolve the long-dependency problem effectively and operate well in parallel, showing good performance on many natural language processing tasks [13]. In [14], the authors proposed a transformer-based network for speech enhancement while it has relatively large model size.

Inspired by the capability of the transformer in sequence modeling, and the effectiveness of the dual-path network for extracting contextual information [15], we here propose a two-stage transformer neural network (TSTNN) for end-to-end monaural speech enhancement in the time domain. The proposed model incorporates the TSTM between the encoder and decoder to learn both local and global contextual information of long-range speech sequences. Our extensive experiments on benchmark dataset show that the TSTNN outperforms the state-of-the-art methods in terms of most evaluation criteria while incurs relatively light model complexity.

## 2. TWO-STAGE TRANSFORMER

In this section, we first present an improved version of the general transformer structure, based on which we propose a two-stage transformer block for extracting local and global contextual information of speech feature.

### 2.1. Improved transformer

The general transformer structure consists of encoder and decoder networks [13]. In our model, we only use the encoder part since the input mixtures and output enhanced sequences have the same length in speech denoising. The original transformer encoder is comprised of three important modules: positional encoding, multi-head attention and position-wise feed-forward network. Different from that, we remove the positional encoding part since it is not suitable for acoustic sequence. Inspired by the effectiveness of RNNs in tracking order information, the first fully connected layer of feed-forward network is replaced with a GRU layer to learn the positional information [16, 17], as shown in Fig. 1.

In multi-head attention block, first, the input ($X$) is mapped with different, learnable linear transformation $h$ times to get queries ($Q$), keys ($K$) and values ($V$) representation, respectively, as described in Eq. (1). Then, the dot product of the query with all keys is computed, followed by division of a constant. After applying the softmax function, the weights on the values are obtained. The attention of each head is the dot product of the weights and values as shown in Eq. (2). The attentions of all heads are concatenated and linearly projected again to obtain the final output in Eq. (3), which is followed by layer normalization and residual connection of input $X$ as given by Eq. (4).

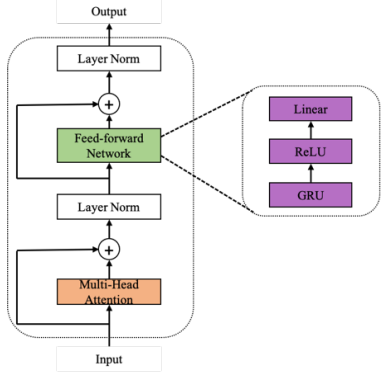

**Fig. 1**: Improved transformer

$$Q_i = XW_i^Q , K_i = XW_i^K , V_i = XW_i^V \quad (1)$$

$$head_i = Attention(Q_i, K_i, V_i) = softmax(\frac{Q_i K_i^T}{\sqrt{d}})V_i \quad (2)$$

$$MultiHead(Q, K, V) = Concat(head_1, \dots, head_h)W^O \quad (3)$$

$$Mid = LayerNorm(X + Multihead) \quad (4)$$

where $X \in \mathbb{R}^{l \times d}$ is the input with sequences of length $l$ and dimension $d$, $i = 1, 2, \dots h$ and $Q_i, K_i, V_i \in \mathbb{R}^{l \times d/h}$ are the mapped queries, keys and values respectively, $W_i^Q, W_i^K, W_i^V \in \mathbb{R}^{d \times d/h}$ denote the $i$th linear transformation matrix for queries, keys and values, respectively. $W^O \in \mathbb{R}^{d \times d}$ is the linear transformation matrix and $h$ is the number of parallel attention layers which is set as 4 in our model.

Then, the output from multi-head attention block is processed by feed-forward network to obtain the final output of improved transformer encoder, where residual connections and layer normalization [18] are add as well. The procedures are defined as follows:

$$FFN(Mid) = ReLU(GRU(Mid))W_1 + b_1 \quad (5)$$

$$Output = LayerNorm(Mid + FFN) \quad (6)$$

where $FFN(\cdot)$ denotes the output of the position-wise feed-forward network, and $W_1 \in \mathbb{R}^{d_{ff} \times d}$, $b_1 \in \mathbb{R}^d$, and $d_{ff} = 4 \times d$.

### 2.2. Two-stage transformer block

We propose a two-stage transformer block based on the improved transformer. As shown in Fig. 2, it has a local transformer and a global transformer, which extracts local and global contextual information, respectively. More specifically, the input is 3-D tensor ($[C, N, F]$), and the local transformer is first applied to individual chunks to parallelly process local information, which performs on the last dimension $F$ of input tensor. Then the global transformer is used for fusing the information of output from local transformer to learn global dependency, which implements on the dimension $N$ of tensor. Besides, each transformer is followed by the group normalization operation and utilizes residual connection.

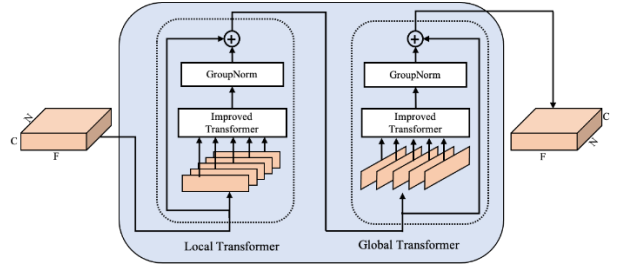

**Fig. 2**: Two-stage transformer block

### 3. PROPOSED MODEL

In this section, we propose a two-stage transformer based neural network (TSTNN) for speech enhancement. As shown in Fig. 3, the new model consists of segmentation stage, encoder, two-stage transformer module, masking module, decoder and overlap-add stage.

### 3.1. Segmentation and overlap-add

The segmentation stage splits raw mixture $X \in \mathbb{R}^{1 \times L}$ into frames of length $F$ with hop size $H$. Then all the frames are stacked to form a 3-D tensor $I \in \mathbb{R}^{1 \times N \times F}$ as the input of encoder. Here $L$ is the length of input mixture, and $N$ denotes the number of frames as given by:

$$N = [(L - F)/(F - H) + 1] \quad (7)$$

The overlap-add method is used as the inverse operation of segmentation, which is used for recovering enhanced waveform.

### 3.2. Encoder

The encoder uses two convolutional layers of which the first one is increasing the number of channels to 64 using convolution with filter of size (1, 1) and the second one halves the dimension of frame size using filter of size (1, 3) with a stride of (1, 2), where a dilated-dense block [19] with four dilation convolution layers is incorporated between them. All convolutional layers are followed by layer normalization and PReLU nonlinearity [20].

### 3.3. Two-stage transformer module (TSTM)

The TSTM consists of four stacked two-stage transformer blocks. Before feeding the output from encoder into TSTM, we halve the channel dimension using convolution with a kernel of size (1, 1) followed by the PReLU nonlinearity, which reduces the computational complexity of the following transformer network. Next, feature representation is processed by TSTM to learn local and global contextual features.

### 3.4. Masking module

Masking network makes use of the output features from TSTM to obtain the mask for denoising. The output from TSTM is doubled along the channel dimension with PReLU nonlinearity and convolution for matching the output of encoder. Then it passes through a two-path 2-D convolution and nonlinearity operation, with the outputs being multiplied together as the input for 2-D convolution and ReLU to get the mask. The final masked encoder feature is obtained by the element-wise multiplication between the mask and the output of the encoder.

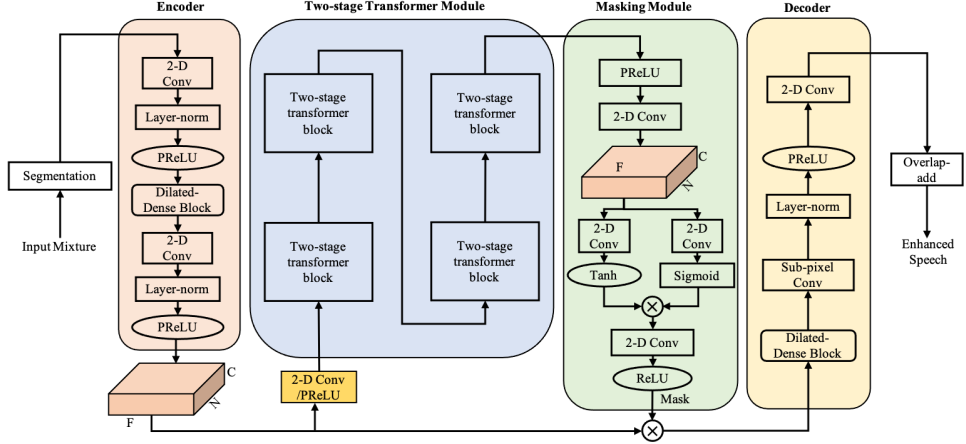

**Fig. 3**: Two-stage transformer neural network (TSTNN) (Note: F, N and C denote frame size, the number of frames and channel, respectively)

### 3.5. Decoder

In the decoder, a dilated dense block and a sub-pixel convolution [21] are applied to reconstruct the masked encoder feature into enhanced speech feature. Then the 2-D convolution with filter size of (1, 1) recovers the channel dimension of the enhanced speech feature into 1 and produces enhanced speech waveform by overlap-add method.

### 3.6. Loss function

Our loss function combines both time domain and time-frequency domain losses. The loss in time-frequency domain can supervise the model to learn more information, leading to higher speech intelligibility and perceptual quality [19], which is defined as:

$$loss_F = \frac{1}{TF}\sum_{t=0}^{T-1}\sum_{f=0}^{F-1}[(|X_r(t,f)| + |X_i(t,f)|) - (|\hat{X}_r(t,f)| + |\hat{X}_i(t,f)|)] \quad (8)$$

where $X$ and $\hat{X}$ denote the spectrogram of clean waveform and the spectrogram of enhanced waveform. $r$ and $i$ are the real and imaginary parts of the complex variable. $T$ and $F$ are the number of time frames and the number of frequency bins, respectively.

The time-domain loss is based on the mean square error (MSE) between the denoised speech and clean speech, which is defined as:

$$loss_T = \frac{1}{N}\sum_{i=0}^{N-1}(x_i - \hat{x}_i)^2 \quad (9)$$

where $x$ and $\hat{x}$ are the sample of the clean speech and the denoised speech, respectively, and $N$ denotes the number of samples.

The final loss function combines these two types of losses mentioned above, as follows:

$$loss = \alpha * loss_F + (1 - \alpha)loss_T \quad (10)$$

where $\alpha$ is a tunable parameter and set as 0.2 in our work.

## 4. EXPERIMENTS

### 4.1. Datasets

We evaluate our proposed model on a standard speech dataset from [22], which is selected from Voice Bank corpus [23], including 11572 utterances of 28 speakers (14 female and 14 male) for training set and 824 utterances of 2 speakers (one male and one female) for testing set. The noisy speech is generated with 10 types of noises (8 from DEMAND dataset [24] and 2 artificially generated) at SNRs of 15 dB, 10 dB, 5 dB and 0 dB for training, and with 5 types of unseen noises at SNRs of 17.5 dB, 12.5 dB, 7.5 dB and 2.5 dB for testing.

### 4.2. Experimental setup

All the utterances are resampled to 16 kHz. Each frame has a size of 512 samples (32ms) with an overlap of 256 samples (16ms). We randomly slice segment of 4 seconds from an utterance if it is larger than 4 seconds. Within a batch, the smaller utterances are zero-padded to match the size of largest utterance.

In the training stage, we train our model for 100 epochs and optimize it by Adam. We use the gradient clipping with maximum L2-norm of 5 to avoid gradient explosion. For learning rate, we use the dynamic strategies during the training stage [13]. More specifically, we first linearly increase the learning rate within $num\_warmups$ training steps, and then decay it by 0.98 for every two epochs as follows:

$$lr = k_1 \cdot d_{model}^{-0.5} \cdot n \cdot num\_warmups^{-1.5}, \quad n \leq num\_warmups$$

$$lr = k_2 \cdot 0.98^{\lfloor \frac{epoch}{2} \rfloor}, \quad n > num\_warmups$$

where $n$ is the number of steps, and $k_1$, $k_2$ are tunable parameters. In our experiments, we set $k_1 = 0.2$, $k_2 = 4e^{-4}$ and $num\_warmups = 4000$. Finally, $d_{model}$ denotes the feature size of the input of transformer which is set as 64 in our paper.

### 4.3. Evaluation metrics

We evaluate the proposed speech enhancement model on several objective measures (Table 1): perceptual evaluation of speech quality (PESQ) [25] with a score range from -0.5 to 4.5; Short-time objective intelligibility (STOI) [26] with a score range from 0 to 1. We also adopt subjective mean opinion scores (MOSs) [27] including CSIG for signal distortion, CBAK for noise distortion evaluation and COVL for overall quality evaluation. All MOSs range from 1 to 5; Segmental signal-to-noise ratio (SSNR) [28] with a value range from -10 to 35 is also used.

### 4.4. Comparison with other time-domain methods

Our proposed model is compared with several other waveform-based methods. As seen from Table 1, TSTNN outperforms, in terms of PESQ score, most existing waveform-based methods and achieves comparable performance with only about 36 times fewer of parameters to DEMUCS with large model configuration (Fig. 4). For STOI value, TSTNN achieves the best score (95%) among all the existing time-domain models. Besides, TSTNN achieves best score in three MOSs evaluation compared with existing time-domain models.

**Table 1**: Evaluation results of our proposed model compared with other methods on the same dataset [22]. Six objective metrics and the number of trainable parameters are considered. (Higher scores are better except parameter size, 'F' denotes frequency and 'T' is time)

| Model | Domain | PESQ | STOI (%) | CSIG | CBAK | COVL | SSNR | Para.(Million) |
|---|---|---|---|---|---|---|---|---|
| Noisy | - | 1.97 | 91 | 3.34 | 2.44 | 2.63 | 1.73 | - |
| Wiener | - | 2.22 | - | 3.23 | 2.68 | 2.67 | 5.07 | - |
| SEGAN, 2017[8] | T | 2.16 | 93 | 3.48 | 2.94 | 2.80 | 7.73 | 97.47 |
| Wavenet, 2018[9] | T | - | - | 3.62 | 3.23 | 2.98 | - | - |
| CNN-GAN, 2018[5] | F | 2.34 | 93 | 3.55 | 2.95 | 2.92 | - | - |
| Wave U-Net, 2018[7] | T | 2.40 | - | 3.52 | 3.24 | 2.96 | 9.97 | 10.0 |
| MMSE-GAN, 2018[2] | F | 2.53 | 93 | 3.80 | 3.12 | 3.14 | - | - |
| MetricGAN, 2019[3] | F | 2.86 | - | 3.99 | 3.18 | 3.42 | - | - |
| DCUnet-16, 2019[6] | F | 2.93 | - | 4.10 | **3.77** | 3.52 | **14.44** | 2.3 |
| DEMUCS (small), 2020[11] | T | 2.93 | 95 | 4.22 | 3.25 | 3.52 | - | 18.9 |
| DEMUCS (large), 2020[11] | T | **3.07** | 95 | 4.31 | 3.4 | 3.63 | - | 33.5 |
| TSTNN (ours) | T | 2.96 | **95** | **4.33** | 3.53 | **3.67** | 9.70 | **0.92** |

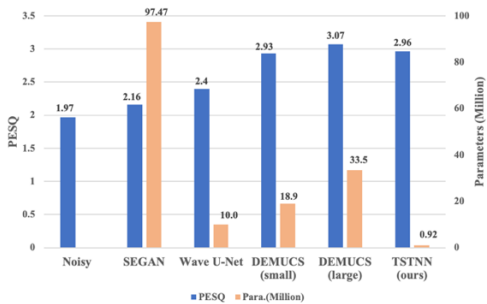

**Fig. 4**: Comparison in terms of PESQ and model size among time-domain methods

### 4.5. Comparison with T-F methods

Table 1 also shows that TSTNN outperforms all T-F methods in most evaluation metrics, especially the PESQ. Moreover, our TSTNN has 0.92 million parameters, which is 2.5 times fewer than DCUNet-16 (2.3 million parameters), achieving extremely low model complexity compared with all other T-F models in Table 1.

### 4.6. The influence of two-stage transformer block

In our model, the two-stage transformer block is designed to extract the features of input speech. In order to further demonstrate the effectiveness of our proposed transformer block, we also designed another architecture for comparison. In this architecture, we use two transformer blocks rather than four blocks in the TSTNN, and we increase the number of encoder layers, which is used as the main feature extractor. Correspondingly, the number of decoder layers is increased to match the encoder layers. In this comparison architecture, we set 4 encoder layers while only one encoder layer in TSTNN. The configuration of each encoder and decoder in the comparison model is the same as the counterparts of TSTNN.

**Table 2**: TSTNN vs. Comparison model.

| Model | PESQ | STOI(%) | CSIG | CBAK | COVL | SSNR |
|---|---|---|---|---|---|---|
| TSTNN | **2.96** | 95 | **4.33** | **3.53** | **3.67** | **9.70** |
| Comparison | 2.87 | 95 | 4.19 | 3.47 | 3.54 | 9.65 |

**Table 3**: Parameters of TSTNN and comparison model.

| Model | TSTNN | Comparison |
|---|---|---|
| Para.(million) | **0.92** | 2.37 |

From Tables 2 and 3, we can see that TSTNN has better scores in all evaluation metrics with 2.6 times fewer parameters than the comparison model. It shows that the transformer block is more efficient than encoder layers in terms of feature extraction. The reason for the remarkable performance could be the properties of our two-stage transformer that it not only works well on long-range sequence but also extracts both local and global contextual information, which outperforms most current architectures.

## 5. CONCLUSION

In this study, we proposed a two-stage transformer neural network for monaural speech enhancement in the time domain, which efficiently extracts both local and global contextual information for long-range speech sequences. Experimental results showed that our model outperforms most of the state-of-the-art methods for most evaluation metrics. Furthermore, our proposed model has much fewer trainable parameters compared with other current models.

## 6. REFERENCES

[1] S.-W. Fu, T.-y. Hu, Y. Tsao, and X. Lu, "Complex spectrogram enhancement by convolutional neural network with multi-metrics learning," in *2017 IEEE 27th International Workshop on Machine Learning for Signal Processing (MLSP)*. IEEE, 2017, pp. 1–6

[2] M. H. Soni, N. Shah, and H. A. Patil, "Time-Frequency Masking-Based Speech Enhancement Using Generative Adversarial Network," in *2018 IEEE International Conference*

*on Acoustics, Speech and Signal Processing (ICASSP)*. IEEE, 2018, pp. 5039–5043.

[3] S.-W. Fu et al., "Metricgan: Generative adversarial networks based black-box metric scores optimization for speech enhancement," in *ICML*, 2019.

[4] N. Shah, H. A. Patil, and M. H. Soni, "Time-Frequency Mask-based Speech Enhancement using Convolutional Generative Adversarial Network," in *Proceedings, APSIPA Annual Summit and Conference*, 2018, vol. 2018, pp. 12–15.

[5] N. Takahashi, P. Agrawal, N. Goswami, and Y. Mitsufuji, "PhaseNet: Discretized Phase Modeling with Deep Neural Networks for Audio Source Separation," in *Proc. Interspeech 2018*, 2018, pp. 2713–2717.

[6] H.-S. Choi, J.-H. Kim, J. Huh, A. Kim, J.-W. Ha, and K. Lee, "Phase-aware Speech Enhancement with Deep Complex U-Net," Mar. 2019.

[7] C. Macartney and T. Weyde, "Improved speech enhancement with the wave-u-net," *arXiv preprint arXiv:1811.11307*, 2018.

[8] Santiago Pascual, Antonio Bonafonte, and Joan Serra, "Segan: Speech enhancement generative adversarial network," *arXiv preprint arXiv:1703.09452*, 2017.

[9] D. Rethage, J. Pons, and X. Serra, "A Wavenet for Speech Denoising," in *2018 IEEE International Conference on Acoustics, Speech and Signal Processing (ICASSP)*, 2018, pp. 5069–5073.

[10] A. Pandey and D. Wang, "TCNN: Temporal convolutional neural network for real-time speech enhancement in the time domain," in *ICASSP*, 2019, pp. 6875–6879

[11] Defossez A, Synnaeve G, Adi Y, "Real Time Speech Enhancement in the Waveform Domain". *arXiv preprint arXiv:2006.12847*, 2020.

[12] Yu F, Koltun V, "Multi-scale context aggregation by dilated convolutions." *arXiv preprint arXiv:1511.07122*, 2015.

[13] A. Vaswani, N. Shazeer, N. Parmar, J. Uszkoreit, L. Jones, A. N. Gomez, Ł. Kaiser, and I. Polosukhin, "Attention is all you need," in *Advances in neural information processing systems*, 2017, pp. 5998–6008.

[14] J. Kim, M. El-Khamy, and J. Lee, "T-gsa: Transformer with gaussian-weighted self-attention for speech enhancement," in *ICASSP 2020-2020 IEEE International Conference on Acoustics, Speech and Signal Processing (ICASSP)*. IEEE, 2020, pp. 6649–6653.

[15] Y. Luo, Z. Chen, and T. Yoshioka, "Dual-path rnn: efficient long sequence modeling for time-domain single-channel speech separation," *in ICASSP 2020-2020 IEEE International Conference on Acoustics, Speech and Signal Processing (ICASSP)*. IEEE, 2020, pp. 46–50.

[16] M. Sperber, J. Niehues, G. Neubig, S. Stuker, and A. Waibel, "Self-attentional acoustic models," *Proc. Interspeech 2018*, pp. 3723–3727, 2018.

[17] J Chen, Q Mao, D. Liu "Dual-Path Transformer Network: Direct Context-Aware Modeling for End-to-End Monaural Speech Separation". *arXiv preprint arXiv:2007.13975*, 2020.

[18] J. L. Ba, J. R. Kiros, and G. E. Hinton, "Layer normalization," *arXiv preprint arXiv:1607.06450*, 2016.

[19] Pandey, A., & Wang, D. "Densely Connected Neural Network with Dilated Convolutions for Real-Time Speech Enhancement in The Time Domain." In *ICASSP 2020-2020 IEEE International Conference on Acoustics, Speech and Signal Processing (ICASSP)*. IEEE, 2020, pp. 6629-6633.

[20] K. He, X. Zhang, S. Ren, and J. Sun, "Delving deep into rectifiers: Surpassing human-level performance on imagenet classification," in *IEEE International Conference on Computer Vision, 2015*, pp. 1026–1034.

[21] W. Shi, J. Caballero, F. Huszar, J. Totz, A. P. Aitken, R. Bishop, D. Rueckert, and Z. Wang, "Real-time single image and video super-resolution using an efficient sub-pixel convolutional neural network," in *IEEE conference on computer vision and pattern recognition*, 2016, pp. 1874–1883.

[22] Valentini-Botinhao C, Wang X, Takaki S, et al. "Investigating RNN-based speech enhancement methods for noise-robust Text-to-Speech." In *SSW*. 2016: 146-152.

[23] Veaux C, Yamagishi J, King S. "The voice bank corpus: Design, collection and data analysis of a large regional accent speech database." In *2013 International Conference Oriental COCOSDA held jointly with 2013 Conference on Asian Spoken Language Research and Evaluation(O-COCOSDA/CASLRE)*. IEEE, 2013: 1-4.

[24] Joachim Thiemann, Nobutaka Ito, and Emmanuel Vincent, "The diverse environments multi-channel acoustic noise database: A database of multichannel environ- mental noise recordings," *The Journal of the Acoustical Society of America*, vol. 133, no. 5, pp. 3591–3591, 2013.

[25] Loizou P C. *Speech enhancement: theory and practice*. CRC press, 2013.

[26] C. H. Taal et al., "An algorithm for intelligibility prediction of time–frequency weighted noisy speech," *IEEE Transactions on Audio, Speech, and Language Processing*, vol. 19, no. 7, pp. 2125–2136, 2011.

[27] Hu Y, Loizou P C. "Evaluation of objective quality measures for speech enhancement." *Audio, Speech, and Language Processing, IEEE Transactions* on, 2008, 16(1): 229-238.

[28] Hansen J H L, Pellom B L. "An effective quality evaluation protocol for speech enhancement algorithms." in *Fifth international conference on spoken language processing*. 1998.